\documentclass[12pt,preprint]{emulateapj}

\usepackage{subfigure}

\newcommand{\msun}{$M_\odot$}

\slugcomment{ApJ 3-Aug-2010}

\shorttitle{Imaging two Herbig Ae Stars with CARMA}
\shortauthors{Hamidouche, M.}

\begin{document}


\title{Aperture Synthesis Imaging of V892 Tau and PV Cep: \\ Disk Evolution}


\author{M. Hamidouche}
\affil{Stratospheric Observatory for Infrared Astronomy\\ NASA Ames Research Center, MS N211-3, Moffett Field, CA 94035}
\email{mhamidouche@sofia.usra.edu}

\begin{abstract}
 I present a study of two Herbig Ae stars that are in completely different evolutionary stages: V892 Tau and PV Cep. Using sub arc-second interferometric observations obtained with the Combined Array for Research in Millimeter-wave Astronomy (CARMA) at $\lambda$=1.3 and 2.7 mm, I have for the first time resolved their disks. I deduce that the 5 Myr old V892 Tau has a low dust opacity index $\beta$=1.1 and a disk mass of $\sim$0.03\msun. These values correspond to the growth of its dust into large up to centimeters size structures. In contrast, the very young (a few $\times$10$^5$ yrs) PV Cep has a quite high opacity index $\beta$=1.75 and a more massive disk 0.8\msun. PV Cep has the youngest resolved disk around any Herbig Ae star. Unlike the youngest T Tauri and Class 0 stars, which contain large and processed grains, the young Herbig Ae star, PV Cep, disk contains ISM-like unprocessed dust. This suggests that PV Cep's dust evolution is slower than T Tauri stars'. I also present high spatial resolution interferometric observations of the PV Cep outflow. The outflow inclination is consistent with the orientation of the known Herbig-Haro flow in that region, HH315.
\end{abstract}

\keywords{Stars: formation, pre-main-sequence, circumstellar matter, Herbig Ae/Be, T Tauri - Techniques: interferometric}

\section{Introduction}

Grain growth is expected to occur during the pre-main-sequence phase of star formation in the circumstellar disks, when the disk is still gas-rich. The dust is processed from ISM-like sub-micron size grains into mm and cm-size structures on their way to form planets \citep{beckwith00}. Interferometric sub-millimetric and millimetric observations are playing a key role in the detection and study of these disks. 

The intermediate-mass pre-main-sequence, Herbig AeBe (HAEBE), stars have a mass ranging roughly between 2 to 8\msun~, with a spectral type range B0 to F5 \citep{her60}. They are classified in two sub-types. Herbig Ae stars that have masses  $<$ 5\msun~ and their spectral type vary between B5 and F5. Herbig Be stars are more massive $>$ 5\msun~ and earlier type stars, B0-B5. They are considered as the progenitors of Vega like stars that are surrounded by debris disks with large solid bodies \citep{backman93} and possibly planets \citep[e.g.][]{wilner02}. HAEBE stars are still not as well studied and understood as their lower mass counterparts T Tauri stars (TTS). Theoretical and observational studies of low-mass pre-main-sequence stars (TTS) have shown that they form via accretion from a circumstellar disk, with an associated outflow \citep{beckwith90,mundy96,rodmann06,andrews09}. 

From spectral energy observations, \citet{hillenbrand92} have shown that the infrared excess observed toward a large sample of HAEBE stars is due to thermal emission from the circumstellar material. Using sub-millimeter single dish observations on the JCMT with the SCUBA array camera, \citet{sandell02} have resolved extended circumstellar emission around two dozen HAEBE stars, and thus suggested the existence of massive disks around these stars. High spatial interferometric radio observations later played an important role by resolving circumstellar disks around Herbig Ae stars \citep[e.g.][]{testi03,mh06,isella07}. These observations have shown that those disks have already processed their dust into larger structures. However, the number of known disks around Herbig Ae stars is still very small to make absolute conclusions on their evolution. I present a study of two more Herbig Ae stars V892 Tau (Elias 3-1) and PV Cep, which are in completely different evolutionary status.

The older source V892 Tau was detected first in the near-IR by \citet{elias78}. V892 Tau is located in the Taurus molecular cloud close to the dark cloud L1495 \citep{Skinner93}. It is of spectral type B8$\pm$3 \citep{hernandez04} and of a mass of 2.8\msun~ \citep{smith05} at a distance of 130 pc \citep{torres09}. From its relatively high X-ray energy \citep{mh08} and low H$\alpha$ emission \citep{manoj06}, V892 Tau must be evolved. Its age is estimated at 5 Myr \citep{manoj06}. \citet{hernandez04} classified it as a main-sequence star and hence without a circumstellar disk. On the other hand, from mid-infrared observations, \citep{monnier08} have resolved it as a close binary (7.7 AU) and have shown that it is surrounded by a circumbinary disk. 

The youngest object of this paper is PV Cep. \citet{natta00} reported its age of 0.3 Myr. Using the \citet{siess00} model, I deduce an age $<$ 1Myr \citep{hlsm09}. It is an embedded star located at the head of the cometary nebula GM 29. It is also located slightly below the birthline \citep{fuente98}. PV Cep has a mass of 3.5\msun~ and is classified as an A5e star \citep{the94}.  In their evolutionary classification of HAEBE stars, \citet{fuente98} ranked it in the least evolved type stars, Type 1. PV Cep shows strong P Cygni profiles in Balmer lines \citep{hernandez04} and strong and variable H$\alpha$ \citep{cohen81}. That is due to a strong wind or an outflow. This is consistent with the youth of this object and the energetic bipolar outflow that was detected toward this star, extended over 2.6 pc in projection \citep{levreault84,reipurth97}. Spectral energy distribution of PV Cep between 3-100 $\mu$m defines it as a Class I young star \citep{stapelfeldt96}. 

So far no interferometric observations have been done toward V892 Tau and PV Cep at 1.3 mm. I present the highest resolution aperture synthesis images of V892 Tau and PV Cep ever done, as well as the first observations of PV Cep at 7 mm. In this study, I report that V892 Tau and PV Cep are surrounded by circumstellar disks and discuss their different disk properties. I focus on their dust properties to probe the disk evolutionary status by modeling the disk spectral energy distribution (SED). Their distinctive disk properties are consistent with their respective ages. The dust evolutionary status of the youngest PV Cep is different than the youngest T Tauri and Class 0 stars. I also present a new interferometric observational evidence of the bipolar outflow driven by PV Cep. 

\section{Observations}

I made interferometric millimeter observations with the Combined Array for Research in Millimeter-wave Astronomy (CARMA\footnote{CARMA has since been combined with the SZA array}) toward these two Herbig Ae stars (Table 1). During my observations period, CARMA was a combined array of 15 antennas, six 10.4 m antennas and nine 6.1 m antennas.

I observed V892 Tau at $\lambda$=1.3 mm in C-array configuration, which corresponds to a maximum baseline of $\sim$300 m. These observations were taken in observing cycles of 20 min, with 17 min on source and 3 min on the phase calibrator. The flux calibrator MWC349 was observed at the beginning of the track. In B-array configuration, I observed the star at $\lambda$=2.7 mm, with a maximum baseline of 810 m. The observations were taken in 7 min cycles, with 5 min on source and 2 min on the phase calibrator. The flux calibrator Uranus was observed at the beginning of the track.

PV Cep was also observed at both $\lambda$=1.3 mm and 2.7 mm wavelengths with CARMA. In C-array configuration, I observed PV Cep at $\lambda$=1.3 mm, with a maximum baseline of $\sim$ 300 m. I used observing cycles of 20 min, with 18 min on source and 2 min on the phase calibrator. The flux calibrator Uranus was observed at the beginning of the track. For $\lambda$=2.7 mm, I used both D-  and B-array configurations corresponding to maximum baselines of 110 m and 790 m respectively. I observed in D-array in 22 min cycles, 17 min on source and 5 min on the phase calibrator, and in 8 min cycles, 5 min on source and 3 min on the phase calibrator in B-array. For both tracks, the flux calibrator MWC349 was observed at the beginning. 

For both sources, I configured the correlator in CARMA to have two bands set at 500 MHz for the continuum and one band at 8 MHz centered at 230.538 GHz, $\lambda$=1.3 mm, and 115.271 GHz, $\lambda$=2.7 mm. These configurations provide a total of 1GHz bandwidth in the continuum. I have FFTed and cleaned the data using the MIRIAD astronomical package \citep{sault95}. For PV Cep mapping at 2.7 mm, I combined B- and D-array observations using robust=-0.5. D-array observations are sensitive to the extended circumstellar structures larger than 6\arcsec. The resulting beam size is 0.88\arcsec$\times$0.73\arcsec~ with a position angle of 22$^o$. In this study, I mainly focus on the dust continuum emission. 

I used the VLA (Very Large Array) of the NRAO\footnote{The National Radio Astronomy Observatory is a facility of the National Science Foundation.} in C-array configuration to observe PV Cep with Q-band receivers ($\lambda$=7mm, 43 GHz). The source was not detected with an rms noise $\sigma$=0.19mJy/beam. An upper limit to the flux density of 4$\sigma$ gives 0.76 mJy. 

\section{Observational Results} 

\textit{V892 Tau - } I present in Table 2 the deduced disk parameters obtained by fitting a two-dimensional elliptical Gaussian from MIRIAD. I estimate the outer radius from the Gaussian R$_g$ = 1.4 FWHM along the major axis of the disk to include 99.9\% of the flux, assuming a circular shape for the circumstellar disk. I have resolved the disk of V892 Tau at $\lambda$=2.7 mm (Figure 1). Figure 2 shows the vector averaged visibility amplitude data binned in annuli around the source. The bin size was chosen to provide approximately a same number of $u,v$ visibility points per bin. The visibility amplitude is plotted as a function of the projected baseline length. Compared to the point source, I see that the disk is resolved at the 3 $\sigma$ level of the amplitude, within error bars. The total flux obtained at $\lambda$ = 2.7 mm is consistent with low-resolution observations of \citet{difrancesco97}. 

I deduce a disk inclination of 59$^o$ and a P.A. = 64$^o$. From mid-infrared imaging of V892 Tau, \citep{monnier08} deduce a similar value 60$^o$ of the disk inclination. I deduce a disk radius of R$_g$ $\simeq$ 100 AU using a distance of 130 pc \citep{torres09}. At $\lambda$=1.3 mm, I find a point source (Figure 1) with a disk radius smaller than about 170 AU. The total flux is quite similar within error bars with single dish observations of V892 Tau \citep{henning98}. This shows that there is a negligeable contribution of an envelope emission in our observations. The flux at these wavelengths has a very low contamination by free-free emission from the ionized gas 10\% and 2\%, at 2.7 and 1.3 mm respectively.

\textit{PV Cep - } These CARMA interferometric observations reveal circumstellar disk emission around PV Cep at both observed wavelengths (Figure 3). The deduced disk parameters are given in Table 2. The peak position fits the optical position of PV Cep \citep{ducourant05} within 0.1\arcsec, which is $\sim$5\% of the beam size. This is consistent with the compact radio emission at 3.6 cm detected by \citet{anglada92}. At $\lambda$=1.3 mm, I resolved the disk (Figure 3). The total flux of 280 mJy is 25\% higher than the peak. It is lower than single dish observations of 360 mJy \footnote{G. Sandell private communication} by only 20\%. Unlike interferometers that filter the surrounding emission, single dish observations include emission from the surrounding material. The low difference in flux with single dish observations, which have much larger beams $>$ 6$''$ indicates that there is an insignificant envelope emission contribution to my interferometric observations. Figure 4 is analogous to Figure 2. It shows that  PV Cep is well resolved at $\lambda$ = 1.3 mm compared to a point source emission. 

At 2.7 mm, the total flux 32 mJy is consistent with low resolution observations of \citet{stapelfeldt96}. Figure 4 shows a scattered emission at the longest baselines, $\sim$250k$\lambda$. I therefore consider the disk only partially resolved. The disk sizes at 1.3 and 2.7 mm wavelengths are consistent. The inclination and the position angle are rather different, but the values at 2.7 mm are uncertain. For the disk modeling in Section 4, I use the values from the resolved disk image at 1.3 mm. 

By combining existing observations of PV Cep at $\lambda$ = 3.6 cm from \citet{anglada92} with the upper limit at $\lambda$=7mm in this work, I constrain the non-thermal emission spectral index $\alpha$ $\lesssim$ 0.73 (where $S_\nu \propto \nu^{\alpha}$). This value is consistent with the value deduced by \citet{stapelfeldt96} $\sim$0.6. A positive value of $\alpha$ means that free-free emission is from the ionized gas in the outflow. I deduce a negligible free-free contribution to the 2.7 and 1.3 mm flux, less than 1.4 mJy and 2.3 mJy respectively.

From the molecular line CO(1$\rightarrow$0) emission, I resolve the inner part of an outflow powered by PV Cep, which coincides with the source position and its disk. The outflow position angle is consistent with the disk inclination (Figure 5). The NW emission corresponds to the blue-shifted part of the outflow at a velocity range of -2.3-0.9 km sec$^{-1}$. The SE emission corresponds to the red-shifted part at higher velocities within the range 4.1-8.5 km sec$^{-1}$ . In my interferometric observations, I only see the closest region ($\sim$ 0.25 pc) around PV Cep, thus, the beginning of the outflow. The corresponding CO spectrum is shown in Figure 6. The emission drops at 2-3 km sec$^{-1}$. This corresponds to the emission from the cloud, which I resolved out. This value is also consistent with the cloud velocity deduced by \citet{cohen81}. Figure 5 shows a S-shape like outflow emission. This can be explained by a rotation of the outflow. This suggests that PV Cep outflow rotates around its parallel axis counter-clock wise, if seen pole-on from its northern side. 

The flow wide opening angles show that the outflow swept the surrounding material of the cloud \citep{reipurth97,arce02}. This is likely why I clearly detect the circumstellar disk around the very young embedded PV Cep star. Using single dish observations, \citet{arce02} mapped a large region (a few pc) surrounding PV Cep. They find a not well collimated outflow in the north of PV Cep. This is consistent with the wide angle of the V-shaped blue shifted part (Figure 5). 

\section{Dust Properties}

The interstellar dust is processed in the flared disk or atmosphere component. The grown and large grains sediment toward the disk midplane immediately while the interstellar micron size dust may take about a Myr to reach the midplane \citep[e.g.][]{natta00}. The disk midplane contains the bulk of the material and this is where the grains grow into larger structures and planetesimals by collisions.

\subsection{Disk Parameters and Model}

For this study, I model the circumstellar disk midplane as an optically thick disk and geometrically thin with a hole in the middle. The disk midplane emission is dominant from the far-infrared into the millimeter region of the SED, or the Rayleigh-Jeans component. Modeling this component constrains the dust properties relevant for this study. I add 20\% calibration uncertainties to the systematic errors for my new 2.7 and 1.3 mm CARMA data. In the sub-millimeter, I use new data from SCUBA at 450 and 850 $\mu$m \citep[][submitted to ApJS.]{swh10}. At $\lambda$=7 mm, I use PV Cep flux as an upper limit, and for V892 Tau, I use the flux value from \citet{difrancesco97}. I also add the existing MIPS\footnote{Spitzer Campaign Id 749 (MIPS004000)} $\lambda$=70$\mu$m observations of V892 Tau. 

I model the disk midplane without an envelope and using a power-law characterization \citep{mundy96,mh06,andrews07a}. This is a simple and convenient model for this study. The disk midplane emission is presented by

\begin{equation}
F_\nu = 2\pi \cos i \int_{R_{in}}^{R_{out}} \{1 - exp [\frac{-\Sigma(R)\kappa_\nu}{\cos i}]\} \times B_\nu[T_d(R)] ~~ \frac{R}{D^2} dR
\end{equation}

The dust opacity is characterized by the power-law $\kappa (\nu)$ = $\kappa_{1200}$($\nu$/1200)$^{-\beta}$, where $\nu$ is the frequency, $\beta$ is the opacity index, and $\kappa_{1200}$ = 0.01 cm$^2$ g$^{-1}$. This includes the gas contribution to the total disk mass with a gas-to-dust mass ratio of 0.01 \citep[e.g.][]{natta00}. Due to the low-spatial resolution of my observations, I do not use a radial profile for the opacity index $\beta$. The disk mass M$_D$ is deduced from the constrained value of M$_D$$\times$$\kappa_{1200}$. The distance to the star is $D$. The disk radius $R$ varies from the disk inner radius, which is the middle hole radius, $R_{in}$ to the disk outer radius $R_{out}$.  The surface density profile is defined as $\Sigma(R)$ = $\Sigma_{1AU}$ $\times$ $R^{-p}$. For the temperature profile in the disk midplane, I use a radial power-law $T(R)$ = T$_{1AU}$ $\times$ $R^{-q}$ characterization, which is estimated by \citet{andrews09}.

The disk outer radius and surface density distribution cannot be constrained from the SED modeling \citep[e.g.][]{mh06,andrews07b}. The spatial resolution provided from these observations is not sufficient (Figure 2 \& 4) to constrain the surface density profile and therefore the outer radius from the disk image and visibility. I fix the outer radius R$_{out}$ = R$_g$, which is the best estimate here \citep[see also][]{ricci10}. I fix the value of $p$=1 that is the value found in the well resolved disks around Herbig Ae stars \citep[e.g.][]{testi03,mh06,isella07}.  Nevertheless, I tested the SED model by varying $R_{out}$ and for p=0.5-1.5, where 1.5 value is the derived value for the solar-nebula surface density distribution \citep{hayashi81}. I found that $p$ values do not significantly affect the two relevant disk parameters values, within error bars, the opacity index $\beta$ and the disk mass M$_{D}$.

 The inner radius is not relevant for this study \citep{testi03}. So I fix its value to the sublimation radius of Herbig Ae stars $R_{in}$=0.3 AU \citep{vink06}. The disk inclination is obtained in the previous section. I keep $T_{1AU}$, the index $q$, the dust opacity index $\beta$, and disk mass $M_{D}$ as free parameters. I deduce the models that fit best the observations using a $\chi^2$ minimization method for the first three parameters spanning a three-dimension cube. Once the best fit 3D grid is found, the disk mass is adjusted until I find the lowest $\chi^2$. I evaluate the models that are close to the minimum $\chi^2$ by taking the models with a likelihood $>$ 0.5. I discuss in the following section the dust opacity and disk mass.

\subsection{Modeling Results and Discussion}

I made well sampled SEDs from the far-infrared to $\lambda$=7 mm. The long wavelengths allow probing the dust properties to centimeter size. I deduced from the disk model the dust opacity index $\beta$. I calculated the millimeter spectral index $\alpha_{mm}$ for both stars at millimeter wavelengths. I used only the interferometric data (1.3, 2.7 and 7 mm) to avoid any over estimation of the surrounding cloud emission, and thus the calculated $\alpha_{mm}$ value is robust. 

When the emission is optically thin, the spectral index is well correlated with the opacity index $\alpha_{mm}$-2=$\beta$ \citep{beckwith91}. An optically thin emission probes directly the dust properties independent of any complexity of the disk morphology \citep{testi03}. I varied the value of $\beta$ in my model from 0 to 2. They correspond to a disk containing only non-processed interstellar like dust, $\beta >$1.7 \citep{weingartner01} or dust completely grown into very large particles (``pebbles''), $\beta$=0 \citep{beckwith91}. An intermediate value $\beta \simeq$1 is consistent with the strong contribution of the grains larger than 1 mm \citep{natta07}. Furthermore, \citet{miyake93} interpret such values $\sim$1 corresponding to dust sizes from 0.1-10 cm. The results of SED modeling are presented in Table 3.

\indent \textit{V892 Tau - } SED is shown in Figure 7. A typical fit model is shown in a solid line. Deprojected visibility profile of this typical fit model in Figure 2 (solid line) demonstrates qualitatively that the model and visibility data are consistent to within errors. I deduce from the best fit models an opacity index $\beta$=1.1$\pm$0.1 and a disk mass of 0.035$\pm$0.015 \msun (Table 3). The corresponding disk-to-star mass ratio $M_D$/$M_*$ = 0.01, or a disk mass of 1\% of the stellar mass. The low value of $\beta$ shows that the disk has passed its initial stage and processed the interstellar dust. This is consistent with the disk low mass, since the large structures' emission cannot be seen due to their small surface area. I do not find any model that fits the observations with an opacity $\beta$=0. Thus, the dust is not completely processed into ``pebbles''. I deduce the millimeter spectral index $\alpha_{mm}$=2.93, and thus the ratio of optically thick to optically thin emission is only a few percent at millimeter wavelengths. The contribution of the inner optically thick region is negligible. This shows that my results are quite robust and the shallow observed slope in the SED is due to processed grains. The disk morphology does not affect the deduced grain properties \citep{testi03}.

\noindent V892 Tau is a relatively old pre-main-sequence star surrounded by an evolved disk. V892 Tau has gone through a large step of evolution. Furthermore, the V892 Tau disk's evolutionary phase is similar to the Herbig Ae star's, MWC 275 (HD 163296). They have the same age, 5 Myr \citep{montesinos09}, and a similar mass $\sim$2.5\msun~ \citep{isella07}. They show a shallow millimeter slope with $\alpha_{mm} \simeq$3 and an optically thin emission with an opacity index $\beta$=1. Nevertheless, MWC 275 has a different spectral type, A3 \citep{blondel06}, and is isolated. In contrast, V892 Tau is an earlier type star (B8) and located in Taurus. The similarity of their disks' evolutionary stage shows that their evolution is not correlated with their spectral type. They also have similar H$\alpha$ EW $\sim$20 $\r{A}$, and thus accretion rate \citep[][and references there in]{manoj06}. Their environment clearly does not affect their disk evolution or accretion process. 

\indent \textit{PV Cep - } The SED is shown in Figure 8. The solid line shows a typical fit disk model. As in Figure 2, the corresponding deprojected visibility profile in Figure 4 (solid line) demonstrates qualitatively that the model and visibility data are consistent to within errors. I deduce from the best fit models $\beta$=1.75 and M$_D$=0.76\msun (Table 3). The corresponding disk-to-star mass ratio $M_D$/$M_*$ = 0.2, or a disk mass of 20\% of the stellar mass, which is a relatively high value compared to older Herbig Ae stars.  The dust is clearly not processed into large structures and is rather similar to the ISM (sub-)micron size dust. It has a mm spectral index $\alpha_{mm}$=3.58. I deduce an optically thick emission contribution of only few percent at millimeter wavelengths. The youth of the star $<$ 1Myr is consistent with the disk evolutionary stage. In fact, PV Cep was classified as an A5 star, but, equivalent to a very young T Tauri star that is coming out of its prestellar envelope \citep{cohen81}. This is the youngest resolved disk around any Herbig Ae star. 

\noindent The star has reached the pre-main-sequence stage without processing its dust. In contrast, the lower mass T Tauri stars appear to have already processed their dust at their youngest age. Compared to PV Cep, interferometric observations of eight young (a few $\times$10$^5$ yrs old) TTS provided much smaller dust opacity indexes $\beta \leq$1 \citep{isella09}. From a study of two dozen TTS of different ages, \citet{ricci10} have found that all the youngest stars $\leq$ 1 Myr have $\beta$ $<$ 1. Similarly, interferometric observations of three Class 0 sources have shown that their grains are already processed and grown at this stage \citep{kwon09}. Additionally, even the massive (10\msun) main-sequence star MWC 297 has a very low-mass disk with evolved grains $\beta$=0.1-0.3 which is presumably remnant from a more massive disk when the star was younger \citep{manoj07}. If the Herbig Ae star, PV Cep, would evolve to a main-sequence star in a similar way as TTS, it should have already processed its dust into large grains. This makes the PV Cep case important and worth further investigations.

I suggest that the disk of PV Cep will evolve and the dust grow into large structures. Since PV Cep is relatively massive $M_*$=3.5\msun~ compared to TTS, its horizontal gravity force, $G M_* m_{d}/r^2$, acting on the orbiting dust, $m_{d}$, at a distance $r$ in the disk atmosphere is stronger. The strong stellar horizontal gravity may hold the grains for a longer time in the PV Cep disk atmosphere than in TTS's, before they sediment toward the midplane \citep{beckwith00}. Nevertheless, while rotating in the disk atmosphere, the dust collides and coagulates \citep{dullemond07}, and thus becomes massive enough to be dragged down toward the midplane by the combined disk and stellar vertical forces \citep{natta07}. At that time, the dust would grow further in the midplane via collisions or local gravitational instability \citep{gw73,youdin02}, and thus show similar dust properties as in the more evolved Herbig Ae disks, such as V892 Tau's and MWC 480's \citep{mh06}. Additionally, the longer duration of the dust evolutionary process in the young PV Cep disk compared to the youngest T Tauri stars' appears to be independent of the disk mass or its gravitational force. In fact, PV Cep disk-to-star mass ratio (20\%) is similar to the average T Tauri stars' ratio.

\section{Conclusion}

I have reported in this paper a study of two intermediate-mass Herbig Ae stars that are very different in their evolutionary stages and were born in two different clouds. I studied the early type B8 and quite evolved 5 Myr old star V892 Tau and the later type A5 and young a few $\times$10$^5$ yr deeply embedded PV Cep. I used new CARMA interferometric observations at 1.3 and 2.7 mm to resolve their disks and look into their dust. I have shown that their dust properties are completely different and this is consistent with the two different stellar evolutionary stages. The main results of this study are:

1) I made the highest resolution interferometric millimeter observations ever made of the disks of V892 Tau and PV Cep, resolving for the first time their disks. They are both optically thin at $\lambda$ = 1.3 mm. 

2) I deduce that V892 Tau is an evolved pre-main-sequence star. It has passed through an important phase of its evolution and has a dust opacity index $\beta$=1.1 and disk mass 0.035\msun, which is about 1\% of the stellar mass. The dust in the disk are grown into large centimeter size structures. I find an interesting resemblance between V892 Tau and another 5 Myr old isolated Herbig Ae star MWC 275. From their low disk mass and opacity index $\beta \simeq$ 1, they seem to evolve in a similar way independent of their spectral types and surrounding environment.

3) I show interferometric observations of the CO(1$\rightarrow$0) molecular line emission toward PV Cep. I see clearly an outflow driven by PV Cep that is consistent with its disk inclination. The outflow inclination is also consistent with the orientation of the giant known Herbig-Haro HH 315 flow in that region. The observations suggest that the outflow is rotating around its parallel axis.

4) PV Cep is a very young pre-main-sequence star that just left its prestellar envelope. It is surrounded by a massive disk of 0.76\msun~, which is relatively massive for a Herbig Ae stars. It is 20\% of the stellar mass. The disk is composed of unprocessed ISM-like dust, corresponding to the opacity index $\beta$=1.75. Unlike Class 0 or young T Tauri disks, the PV Cep disk does not contain large structures or grown grains. I suggest that due to the higher mass of PV Cep compared to T Tauri stars', the grains in the PV Cep disk atmosphere will take a longer timescale to grow and settle downward toward the disk midplane. At that time, they will presumably continue their evolution like the more evolved Herbig Ae stars, such as V892 Tau. The PV Cep star is the youngest and the least evolved resolved disk known around any Herbig Ae star. Its dust evolutionary stage is different than the youngest T Tauri and Class 0 stars'.

This raises the question of whether the evolutionary process of Herbig Ae stars is different than T Tauri stars'. In this study, PV Cep case indicates that it is different, or slower, in the beginning, at the age of a few $\times$ 10$^5$ yrs. Although, a study of more disks around more Herbig stars of different ages is important to further probe the evolution of Herbig Ae compared to T Tauri stars' \citep[][in preparation.]{hlm10}. Higher spatial-resolution interferometric observations are necessary for such a study. The new generation of interferometers ALMA and EVLA will provide sensitive high-spatial resolution images. This will allow us to extend the sample of resolved Herbig AeBe disks into the fainter and farther objects, such as the massive Herbig Be stars. 

\acknowledgments
I thank Dana Backman for his comments on disk evolution and for his critical reading of the paper. I thank  G\"oran Sandell for his helpful comments and for our discussions on Herbig stars. I thank Leslie Looney and Pamela Marcum for useful discussions. Support for CARMA construction was derived from the states of California, Illinois, and Maryland, the Gordon and Betty Moore Foundation, the Eileen and Kenneth Norris Foundation, the Caltech Associates, and the National Science Foundation. Ongoing CARMA development
and operations are supported by the National
Science Foundation under a cooperative agreement, and
by the CARMA partner universities.



\begin{thebibliography}{55}
\expandafter\ifx\csname natexlab\endcsname\relax\def\natexlab#1{#1}\fi
\expandafter\ifx\csname url\endcsname\relax
  \def\url#1{\texttt{#1}}\fi
\expandafter\ifx\csname urlprefix\endcsname\relax\def\urlprefix{URL }\fi
\providecommand{\eprint}[2][]{\url{#2}}


\bibitem[{{Andrews} \& {Williams}(2007{\natexlab{a}})}]{andrews07a}
{Andrews}, S.~M. \& {Williams}, J.~P. 2007{\natexlab{a}}, \apj, 671, 1800

\bibitem[{{Andrews} \& {Williams}(2007{\natexlab{b}})}]{andrews07b}
---. 2007{\natexlab{b}}, \apj, 659, 705

\bibitem[{{Andrews} {et~al.}(2009){Andrews}, {Wilner}, {Hughes}, {Qi}, \&
  {Dullemond}}]{andrews09}
{Andrews}, S.~M., {Wilner}, D.~J., {Hughes}, A.~M., {Qi}, C., \& {Dullemond},
  C.~P. 2009, \apj, 700, 1502

\bibitem[{{Anglada} {et~al.}(1992){Anglada}, {Rodriguez}, {Canto}, {Estalella},
  \& {Torrelles}}]{anglada92}
{Anglada}, G., {Rodriguez}, L.~F., {Canto}, J., {Estalella}, R., \&
  {Torrelles}, J.~M. 1992, \apj, 395, 494

\bibitem[{{Arce} \& {Goodman}(2002)}]{arce02}
{Arce}, H.~G. \& {Goodman}, A.~A. 2002, \apj, 575, 911

\bibitem[{{Backman} \& {Paresce}(1993)}]{backman93}
{Backman}, D.~E. \& {Paresce}, F. 1993, in Protostars and Planets III, ed.
  {E.~H.~Levy \& J.~I.~Lunine}, 1253--1304

\bibitem[{{Beckwith} {et~al.}(2000){Beckwith}, {Henning}, \&
  {Nakagawa}}]{beckwith00}
{Beckwith}, S.~V.~W., {Henning}, T., \& {Nakagawa}, Y. 2000, Protostars and
  Planets IV, 533

\bibitem[{{Beckwith} \& {Sargent}(1991)}]{beckwith91}
{Beckwith}, S.~V.~W. \& {Sargent}, A.~I. 1991, \apj, 381, 250

\bibitem[{{Beckwith} {et~al.}(1990){Beckwith}, {Sargent}, {Chini}, \&
  {Guesten}}]{beckwith90}
{Beckwith}, S.~V.~W., {Sargent}, A.~I., {Chini}, R.~S., \& {Guesten}, R. 1990,
  \aj, 99, 924

\bibitem[{{Blondel} \& {Djie}(2006)}]{blondel06}
{Blondel}, P.~F.~C. \& {Djie}, H.~R.~E.~T.~A. 2006, \aap, 456, 1045

\bibitem[{{Cohen} {et~al.}(1981){Cohen}, {Kuhi}, {Spinrad}, \&
  {Harlan}}]{cohen81}
{Cohen}, M., {Kuhi}, L.~V., {Spinrad}, H., \& {Harlan}, E.~A. 1981, \apj, 245,
  920

\bibitem[{{di Francesco} {et~al.}(1997){di Francesco}, {Evans}, {Harvey},
  {Mundy}, {Guilloteau}, \& {Chandler}}]{difrancesco97}
{di Francesco}, J., {Evans}, II, N.~J., {Harvey}, P.~M., {Mundy}, L.~G.,
  {Guilloteau}, S., \& {Chandler}, C.~J. 1997, \apj, 482, 433

\bibitem[{{Ducourant} {et~al.}(2005){Ducourant}, {Teixeira}, {P{\'e}ri{\'e}},
  {Lecampion}, {Guibert}, \& {Sartori}}]{ducourant05}
{Ducourant}, C., {Teixeira}, R., {P{\'e}ri{\'e}}, J.~P., {Lecampion}, J.~F.,
  {Guibert}, J., \& {Sartori}, M.~J. 2005, \aap, 438, 769

\bibitem[{{Dullemond} {et~al.}(2007){Dullemond}, {Hollenbach}, {Kamp}, \&
  {D'Alessio}}]{dullemond07}
{Dullemond}, C.~P., {Hollenbach}, D., {Kamp}, I., \& {D'Alessio}, P. 2007,
  Protostars and Planets V, 555

\bibitem[{{Elias}(1978)}]{elias78}
{Elias}, J.~H. 1978, \apj, 224, 857

\bibitem[{{Fuente} {et~al.}(1998){Fuente}, {Martin-Pintado}, {Bachiller},
  {Neri}, \& {Palla}}]{fuente98}
{Fuente}, A., {Martin-Pintado}, J., {Bachiller}, R., {Neri}, R., \& {Palla}, F.
  1998, \aap, 431, 307

\bibitem[{{Goldreich} \& {Ward}(1973)}]{gw73}
{Goldreich}, P. \& {Ward}, W.~R. 1973, \apj, 183, 1051

\bibitem[{{Hamidouche} {et~al.}(2009){Hamidouche}, {Looney}, {Sandell}, \&
  {Mundy}}]{hlsm09}
{Hamidouche}, M., {Looney}, L., {Sandell}, G., \& {Mundy}, L. 2009, in Bulletin
  of the American Astronomical Society, Vol.~41, Bulletin of the American
  Astronomical Society, 210--+

\bibitem[{{Hamidouche} {et~al.}(2006){Hamidouche}, {Looney}, \& {Mundy}}]{mh06}
{Hamidouche}, M., {Looney}, L.~W., \& {Mundy}, L.~G. 2006, \apj, 651, 321

\bibitem[{{Hamidouche} {et~al.}(2008){Hamidouche}, {Wang}, \& {Looney}}]{mh08}
{Hamidouche}, M., {Wang}, S., \& {Looney}, L.~W. 2008, \aj, 135, 1474

\bibitem[{{{Hamidouche}, M., and {Looney}, L.~W., and {Mundy},
  L.~G.}(2010)}]{hlm10}
{{Hamidouche}, M., and {Looney}, L.~W., and {Mundy}, L.~G.} 2010

\bibitem[{{Hayashi}(1981)}]{hayashi81}
{Hayashi}, C. 1981, Progress of Theoretical Physics Supplement, 70, 35

\bibitem[{{Henning} {et~al.}(1998){Henning}, {Burkert}, {Launhardt}, {Leinert},
  \& {Stecklum}}]{henning98}
{Henning}, T., {Burkert}, A., {Launhardt}, R., {Leinert}, C., \& {Stecklum}, B.
  1998, \aap, 336, 565

\bibitem[{{Herbig}(1960)}]{her60}
{Herbig}, G.~H. 1960, \apjs, 4, 337

\bibitem[{{Hern{\'a}ndez} {et~al.}(2004){Hern{\'a}ndez}, {Calvet},
  {Brice{\~n}o}, {Hartmann}, \& {Berlind}}]{hernandez04}
{Hern{\'a}ndez}, J., {Calvet}, N., {Brice{\~n}o}, C., {Hartmann}, L., \&
  {Berlind}, P. 2004, \aj, 127, 1682

\bibitem[{{Hillenbrand} {et~al.}(1992){Hillenbrand}, {Strom}, {Vrba}, \&
  {Keene}}]{hillenbrand92}
{Hillenbrand}, L.~A., {Strom}, S.~E., {Vrba}, F.~J., \& {Keene}, J. 1992, \apj,
  397, 613

\bibitem[{{Isella} {et~al.}(2009){Isella}, {Carpenter}, \&
  {Sargent}}]{isella09}
{Isella}, A., {Carpenter}, J.~M., \& {Sargent}, A.~I. 2009, \apj, 701, 260

\bibitem[{{Isella} {et~al.}(2007){Isella}, {Testi}, {Natta}, {Neri}, {Wilner},
  \& {Qi}}]{isella07}
{Isella}, A., {Testi}, L., {Natta}, A., {Neri}, R., {Wilner}, D., \& {Qi}, C.
  2007, ArXiv e-prints, 704

\bibitem[{{Kwon} {et~al.}(2009){Kwon}, {Looney}, {Mundy}, {Chiang}, \&
  {Kemball}}]{kwon09}
{Kwon}, W., {Looney}, L.~W., {Mundy}, L.~G., {Chiang}, H., \& {Kemball}, A.~J.
  2009, in , 841--852

\bibitem[{{Levreault}(1984)}]{levreault84}
{Levreault}, R.~M. 1984, \apj, 277, 634

\bibitem[{{Manoj} {et~al.}(2006){Manoj}, {Bhatt}, {Maheswar}, \&
  {Muneer}}]{manoj06}
{Manoj}, P., {Bhatt}, H.~C., {Maheswar}, G., \& {Muneer}, S. 2006, \apj, 653,
  657

\bibitem[{{Manoj} {et~al.}(2007){Manoj}, {Ho}, {Ohashi}, {Zhang}, {Hasegawa},
  {Chen}, {Bhatt}, \& {Ashok}}]{manoj07}
{Manoj}, P., {Ho}, P.~T.~P., {Ohashi}, N., {Zhang}, Q., {Hasegawa}, T., {Chen},
  H., {Bhatt}, H.~C., \& {Ashok}, N.~M. 2007, \apjl, 667, L187

\bibitem[{{Miyake} \& {Nakagawa}(1993)}]{miyake93}
{Miyake}, K. \& {Nakagawa}, Y. 1993, Icarus, 106, 20

\bibitem[{{Monnier} {et~al.}(2008){Monnier}, {Tannirkulam}, {Tuthill},
  {Ireland}, {Cohen}, {Danchi}, \& {Baron}}]{monnier08}
{Monnier}, J.~D., {Tannirkulam}, A., {Tuthill}, P.~G., {Ireland}, M., {Cohen},
  R., {Danchi}, W.~C., \& {Baron}, F. 2008, \apjl, 681, L97

\bibitem[{{Montesinos} {et~al.}(2009){Montesinos}, {Eiroa}, {Mora}, \&
  {Mer{\'{\i}}n}}]{montesinos09}
{Montesinos}, B., {Eiroa}, C., {Mora}, A., \& {Mer{\'{\i}}n}, B. 2009, \aap,
  495, 901

\bibitem[{{Mundy} {et~al.}(1996){Mundy}, {Looney}, {Erickson}, {Grossman},
  {Welch}, {Forster}, {Wright}, {Plambeck}, {Lugten}, \& {Thornton}}]{mundy96}
{Mundy}, L.~G., {Looney}, L.~W., {Erickson}, W., {Grossman}, A., {Welch},
  W.~J., {Forster}, J.~R., {Wright}, M.~C.~H., {Plambeck}, R.~L., {Lugten}, J.,
  \& {Thornton}, D.~D. 1996, \apjl, 464, L169+

\bibitem[{{Natta} {et~al.}(2000){Natta}, {Grinin}, \& {Mannings}}]{natta00}
{Natta}, A., {Grinin}, V., \& {Mannings}, V. 2000, Protostars and Planets IV,
  559

\bibitem[{{Natta} {et~al.}(2007){Natta}, {Testi}, {Calvet}, {Henning},
  {Waters}, \& {Wilner}}]{natta07}
{Natta}, A., {Testi}, L., {Calvet}, N., {Henning}, T., {Waters}, R., \&
  {Wilner}, D. 2007, in Protostars and Planets V, ed. {B.~Reipurth, D.~Jewitt,
  \& K.~Keil}, 767--781

\bibitem[{{Reipurth} {et~al.}(1997){Reipurth}, {Bally}, \&
  {Devine}}]{reipurth97}
{Reipurth}, B., {Bally}, J., \& {Devine}, D. 1997, \aj, 114, 2708

\bibitem[{{Ricci} {et~al.}(2010){Ricci}, {Testi}, {Natta}, {Neri}, {Cabrit}, \&
  {Herczeg}}]{ricci10}
{Ricci}, L., {Testi}, L., {Natta}, A., {Neri}, R., {Cabrit}, S., \& {Herczeg},
  G.~J. 2010, \aap, 512, A15+

\bibitem[{{Rodmann} {et~al.}(2006){Rodmann}, {Henning}, {Chandler}, {Mundy}, \&
  {Wilner}}]{rodmann06}
{Rodmann}, J., {Henning}, T., {Chandler}, C.~J., {Mundy}, L.~G., \& {Wilner},
  D.~J. 2006, \aap, 446, 211

\bibitem[{{Sandell} \& {Weintraub}(2002)}]{sandell02}
{Sandell}, G. \& {Weintraub}, D.~A. 2002, in Bulletin of the American
  Astronomical Society, Vol.~34, Bulletin of the American Astronomical Society,
  763--+

\bibitem[{{{Sandell}, G., and {Weintraub}, D.~A., and {Hamidouche},
  M.}(2010)}]{swh10}
{{Sandell}, G., and {Weintraub}, D.~A., and {Hamidouche}, M.} 2010

\bibitem[{{Sault} {et~al.}(1995){Sault}, {Teuben}, \& {Wright}}]{sault95}
{Sault}, R.~J., {Teuben}, P.~J., \& {Wright}, M.~C.~H. 1995, in Astronomical
  Society of the Pacific Conference Series, Vol.~77, Astronomical Data Analysis
  Software and Systems IV, ed. {R.~A.~Shaw, H.~E.~Payne, \& J.~J.~E.~Hayes},
  433--+

\bibitem[{{Siess} {et~al.}(2000){Siess}, {Dufour}, \& {Forestini}}]{siess00}
{Siess}, L., {Dufour}, E., \& {Forestini}, M. 2000, \aap, 358, 593

\bibitem[{{Skinner} {et~al.}(1993){Skinner}, {Brown}, \& {Stewart}}]{Skinner93}
{Skinner}, S.~L., {Brown}, A., \& {Stewart}, R.~T. 1993, \apjs, 87, 217

\bibitem[{{Smith} {et~al.}(2005){Smith}, {Balega}, {Duschl}, {Hofmann},
  {Lachaume}, {Preibisch}, {Schertl}, \& {Weigelt}}]{smith05}
{Smith}, K.~W., {Balega}, Y.~Y., {Duschl}, W.~J., {Hofmann}, K.-H., {Lachaume},
  R., {Preibisch}, T., {Schertl}, D., \& {Weigelt}, G. 2005, \aap, 431, 307

\bibitem[{{Stapelfeldt}(1996)}]{stapelfeldt96}
{Stapelfeldt}, K. 1996, in IAU Symposium, Vol. 170, CO: Twenty-Five Years of
  Millimeter-Wave Spectroscopy, 104P--+

\bibitem[{{Testi} {et~al.}(2003){Testi}, {Natta}, {Shepherd}, \&
  {Wilner}}]{testi03}
{Testi}, L., {Natta}, A., {Shepherd}, D.~S., \& {Wilner}, D.~J. 2003, \aap,
  403, 323

\bibitem[{{Th\'e} {et~al.}(1994){Th\'e}, {de Winter}, \& {Perez}}]{the94}
{Th\'e}, P.~S., {de Winter}, D., \& {Perez}, M.~R. 1994, \aaps, 104, 315

\bibitem[{{Torres} {et~al.}(2009){Torres}, {Loinard}, {Mioduszewski}, \&
  {Rodr{\'{\i}}guez}}]{torres09}
{Torres}, R.~M., {Loinard}, L., {Mioduszewski}, A.~J., \& {Rodr{\'{\i}}guez},
  L.~F. 2009, \apj, 698, 242

\bibitem[{{Vinkovi{\'c}}(2006)}]{vink06}
{Vinkovi{\'c}}, D. 2006, \apj, 651, 906

\bibitem[{{Weingartner} \& {Draine}(2001)}]{weingartner01}
{Weingartner}, J.~C. \& {Draine}, B.~T. 2001, \apj, 548, 296

\bibitem[{{Wilner} {et~al.}(2002){Wilner}, {Holman}, {Kuchner}, \&
  {Ho}}]{wilner02}
{Wilner}, D.~J., {Holman}, M.~J., {Kuchner}, M.~J., \& {Ho}, P.~T.~P. 2002,
  \apjl, 569, L115

\bibitem[{{Youdin} \& {Shu}(2002)}]{youdin02}
{Youdin}, A.~N. \& {Shu}, F.~H. 2002, \apj, 580, 494

\end{thebibliography}

\clearpage

\begin{figure}
  \begin{center}
    \mbox{
      \subfigure[]{\includegraphics[angle=-90,scale=0.48]{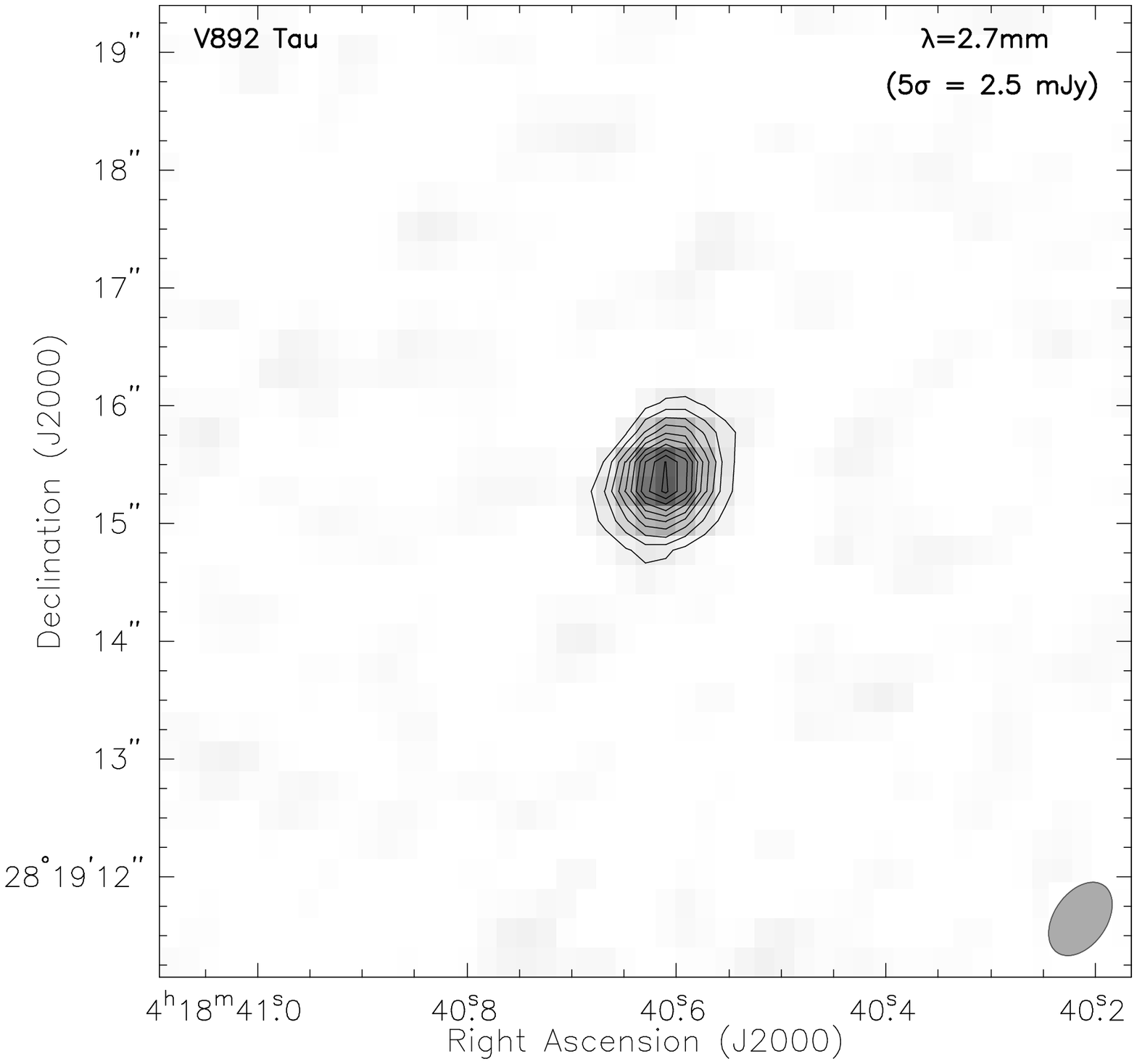}}\quad
      \subfigure[]{\includegraphics[angle=-90,scale=0.48]{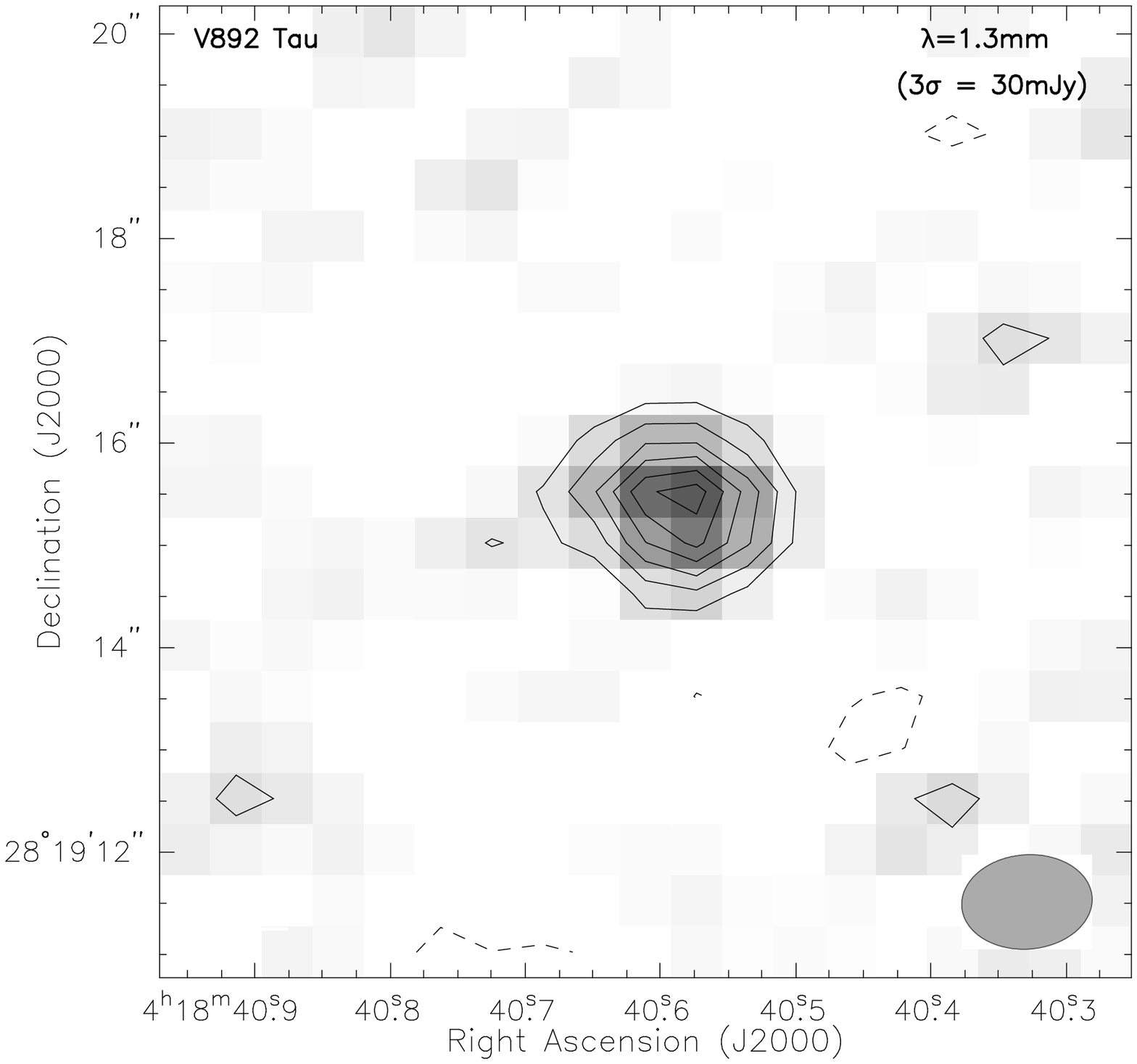}}
      }
    \caption{CARMA images of V892 Tau. The left image is at $\lambda$=2.7 mm and the right at $\lambda$=1.3 mm. The contour steps and the starting levels are shown in each image.\label{mapsElias}}
  \end{center}
\end{figure}

\clearpage

\begin{figure}
  \begin{center}
\includegraphics[angle=0,scale=0.5]{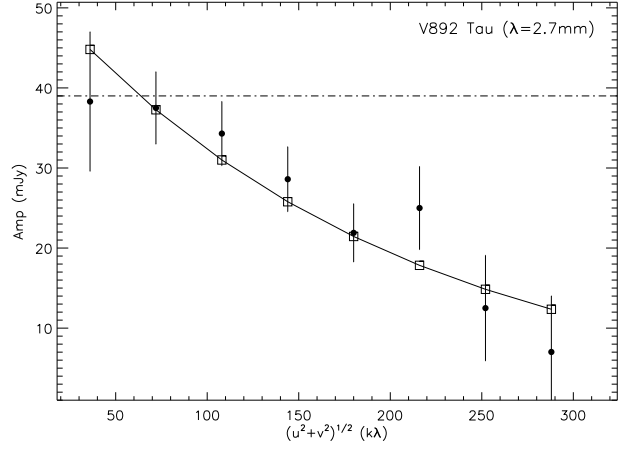}
  \end{center}
\caption{V892 Tau - Vector averaged visibility amplitude as a function of projected baseline length at $\lambda$=2.7 mm. The error bars are standard deviation of visibilities in each bin. This $u,v$ amplitude decreases with the baseline length down to 3 $\sigma$ amplitude level within error bars. This suggests that the disk is resolved. The \textit{dashed line} corresponds to a point source emission model and is shown for comparison with the disk emission. The \textit{solid line (with open squares)} shows deprojected visibility profile from the model (Section 4.2). This is a qualitative demonstration that the model and visibility data are consistent to within errors.}
\end{figure}

\clearpage
\begin{figure}
  \begin{center}
    \mbox{
      \subfigure[]{\includegraphics[angle=-90,scale=0.48]{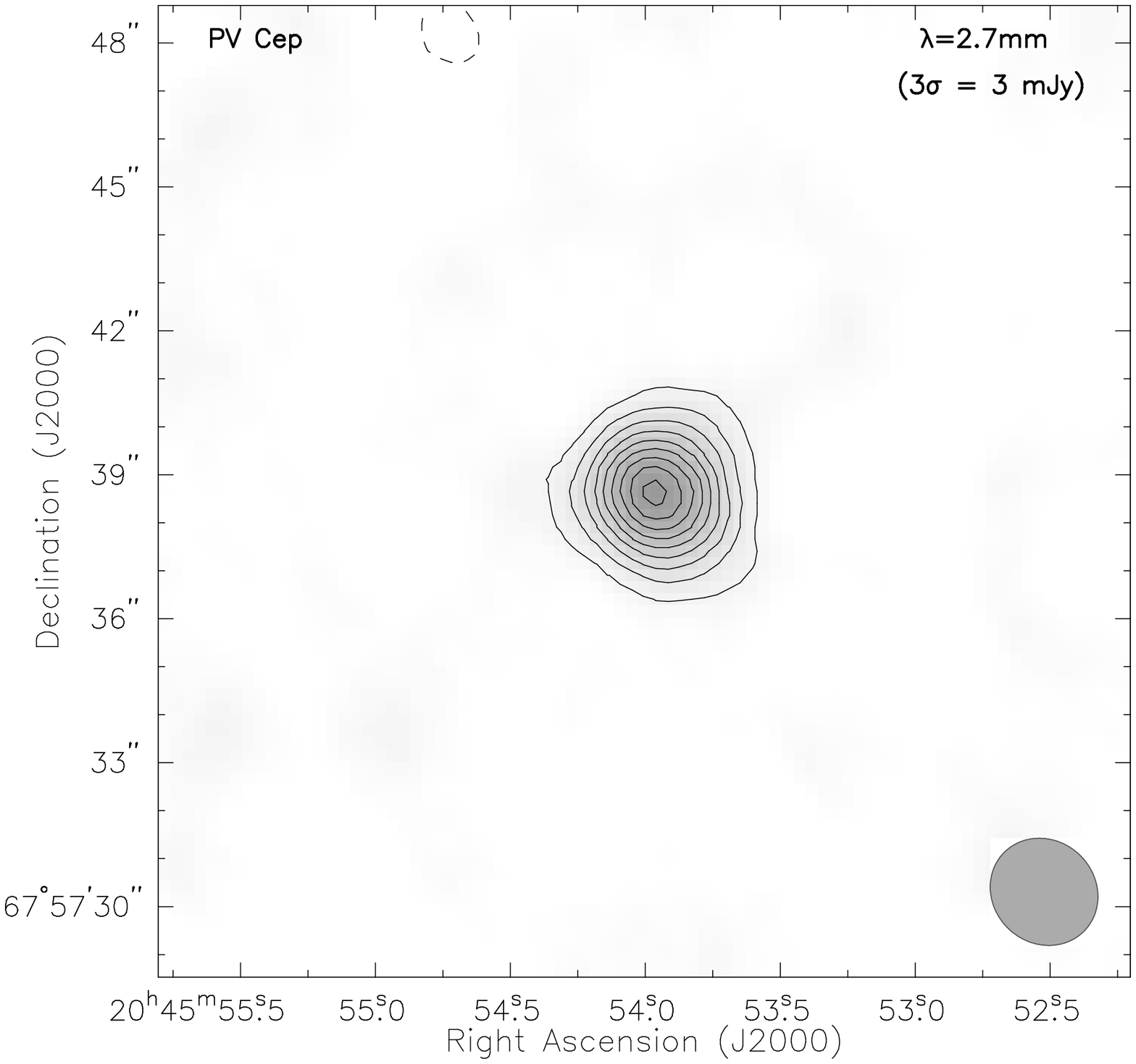}}\quad
      \subfigure[]{\includegraphics[angle=-90,scale=0.48]{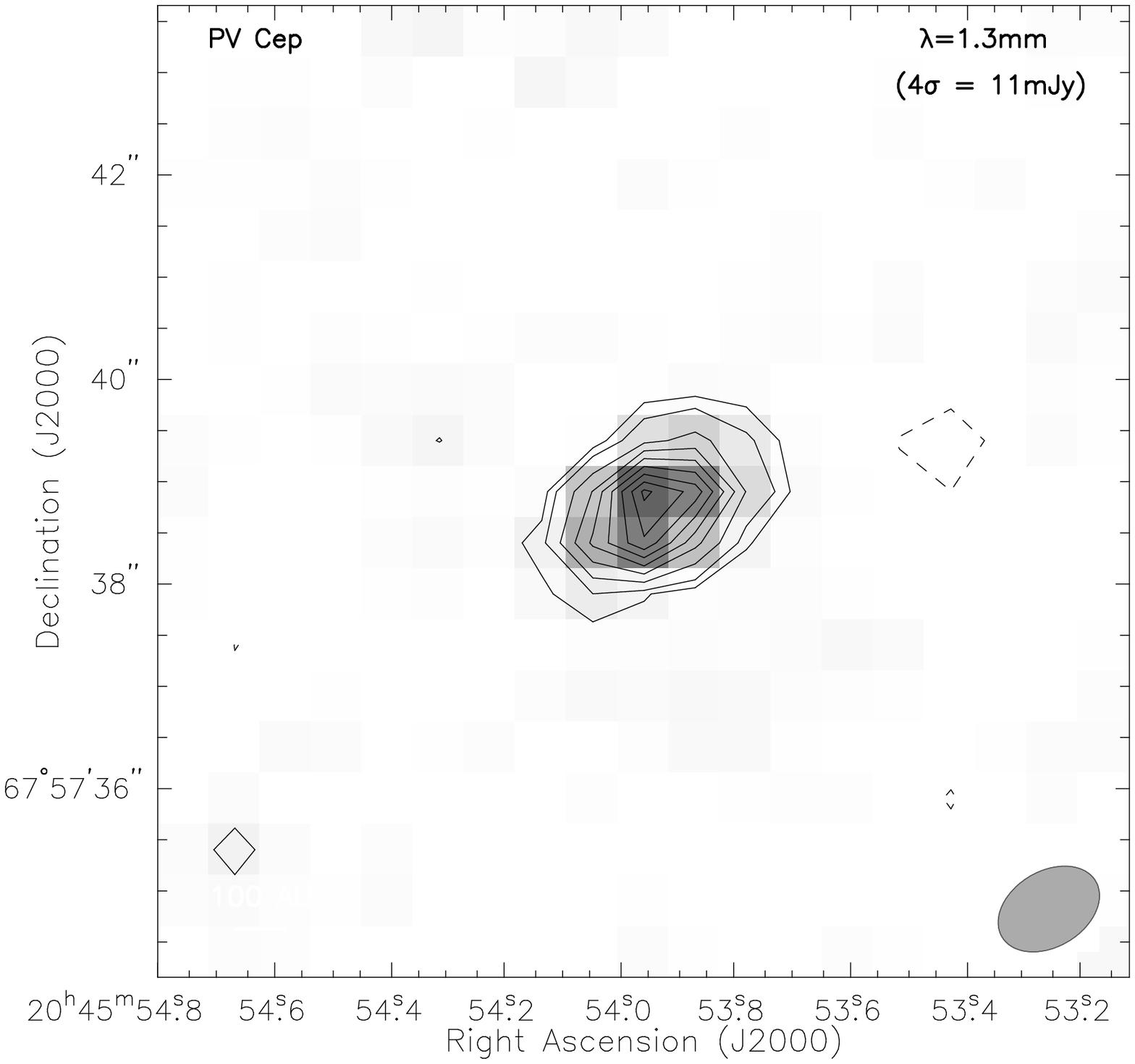}}
      }
    \caption{CARMA images of PV Cep. The left image is at $\lambda$=2.7 mm and the right at $\lambda$=1.3 mm. The contour steps and the starting levels are shown in each image.\label{mapsPvcep}}
  \end{center}
\end{figure}

\clearpage

\begin{figure}
  \begin{center}
    \mbox{
      \subfigure[]{\includegraphics[angle=0,scale=0.48]{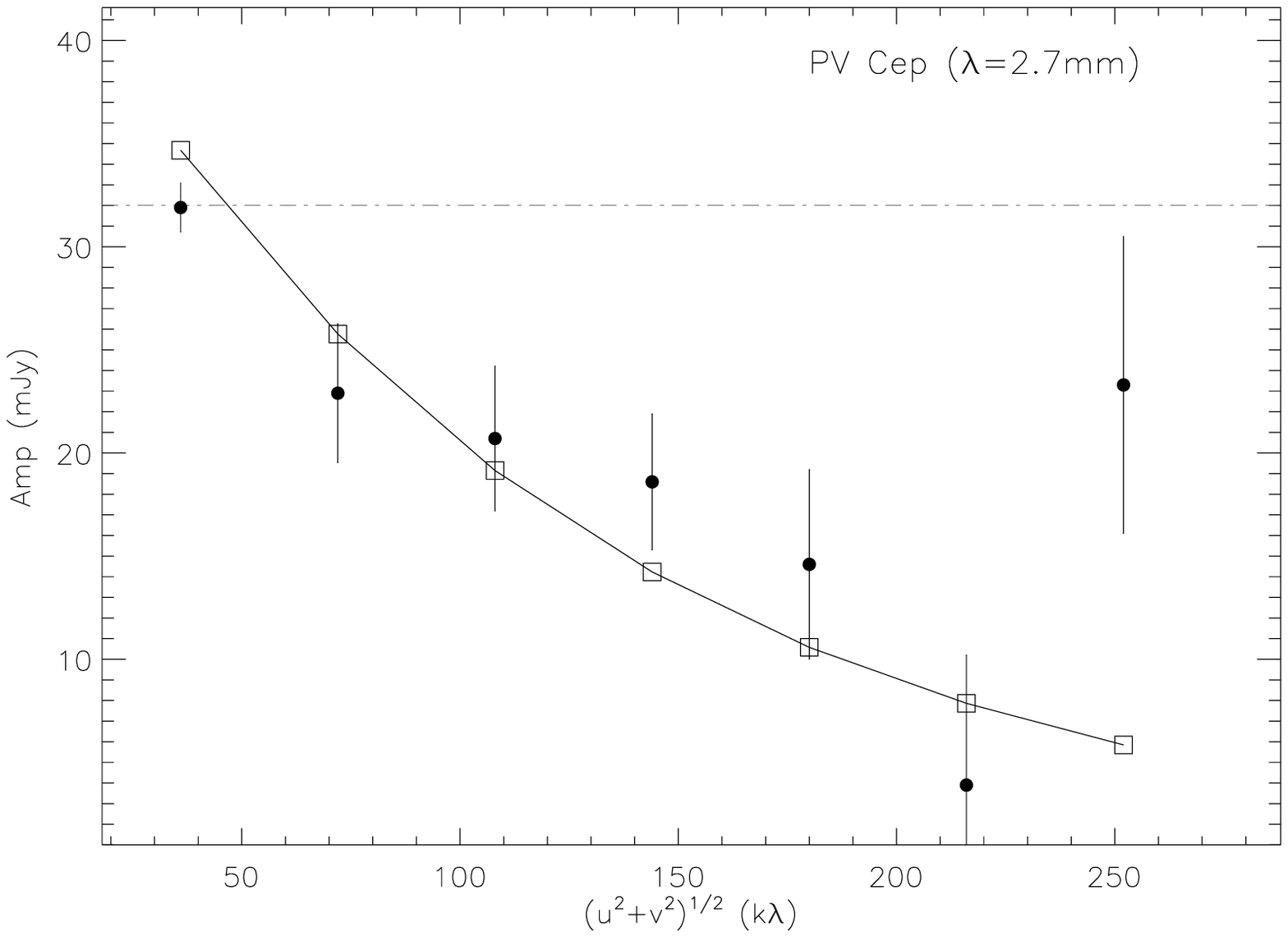}}\quad
      \subfigure[]{\includegraphics[angle=0,scale=0.48]{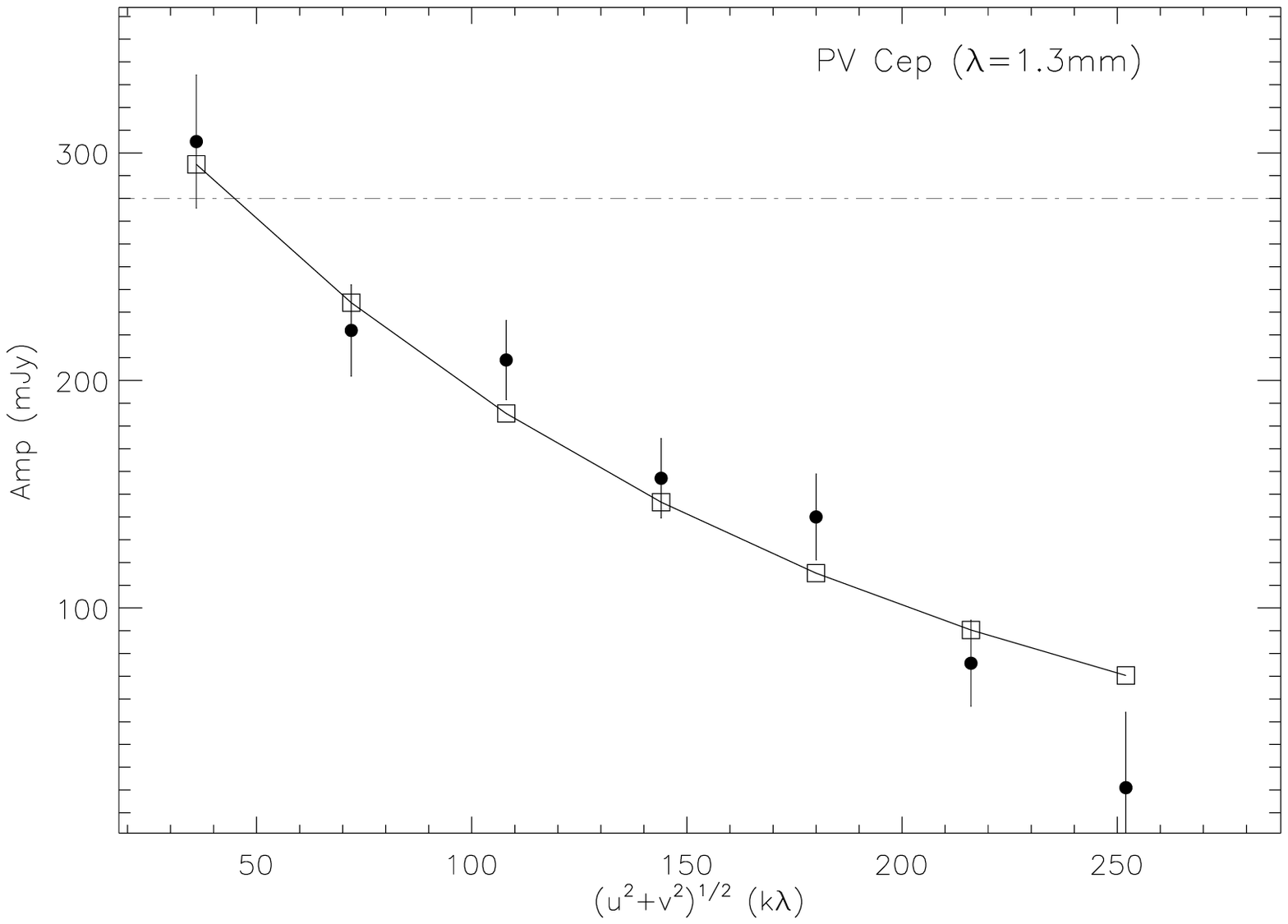}}
      }
  \end{center}
\caption{PV Cep - Vector averaged visibility amplitude as a function of projected baseline length. The error bars correspond to the standard deviation of visibilities in each bin. \textit{(a)}  $\lambda$ = 2.7 mm - The disk is not a point source. But, because of the decoherence at the longest baselines, 250k$\lambda$, I do not consider the disk fully resolved. \textit{(b)} $\lambda$ = 1.3 mm - The data amplitude decreases with the baseline length down to 3 $\sigma$ amplitude level within error bars. This suggests that the disk is resolved. The \textit{dashed lines} correspond to a point source emission model and are shown for comparison with the disk emission. Similar to Fig. 2, for qualitative demonstration, deprojected visibility profiles are plotted with \textit{solid lines (with open squares)} in both figures, showing consistency of model and data visibilities to within the errors (Section 4.2).}
\end{figure}

\clearpage

\begin{figure}
\begin{center}
\includegraphics[angle=-90,scale=0.7]{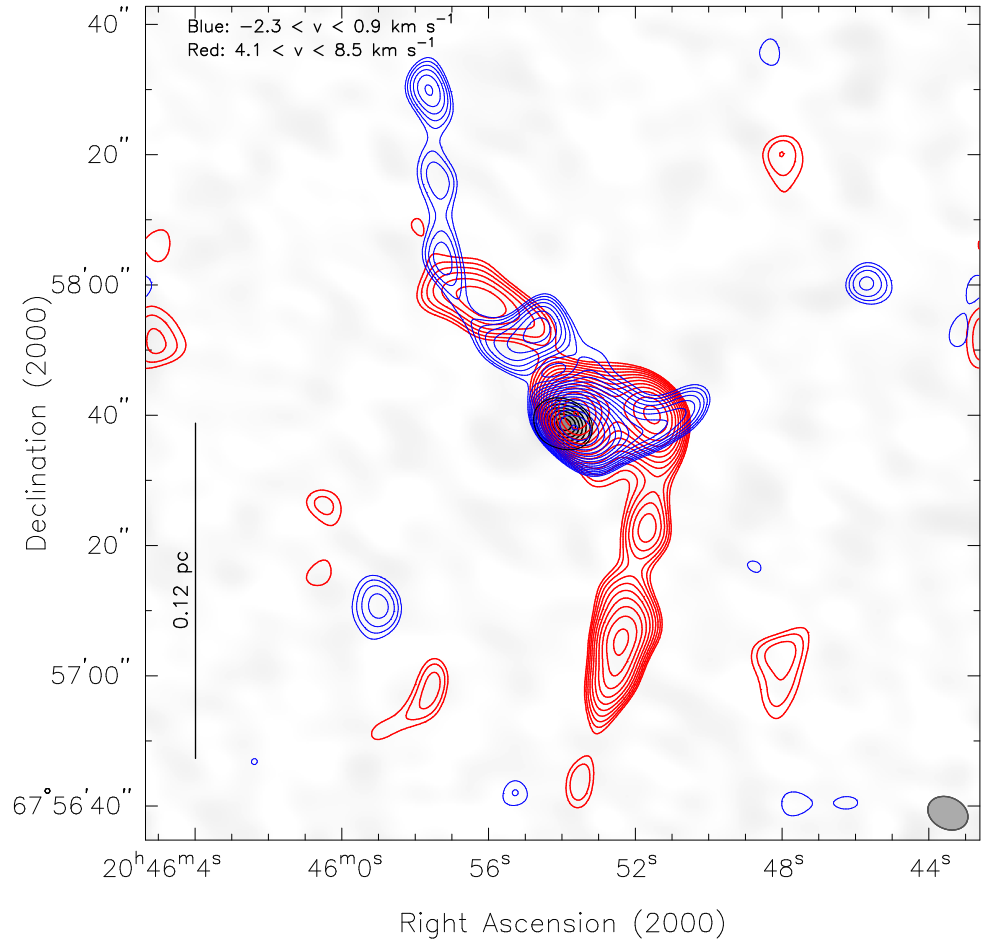}
\end{center}
\caption{PV Cep CO(1$\rightarrow$0) molecular line emission map. The outflow emission is shown in contours for the blue- and red-shifted parts. The ellipse marks the disk position. This emission is obtained after subtracting the continuum emission and convolving the image by a 8'' beam. The contours are in steps of 0.12 and 0.14 K $\times$ km sec$^{-1}$ for the red and the blue respectively.}
\end{figure}

\clearpage
\begin{figure}
\begin{center}
\includegraphics[angle=90,scale=0.5]{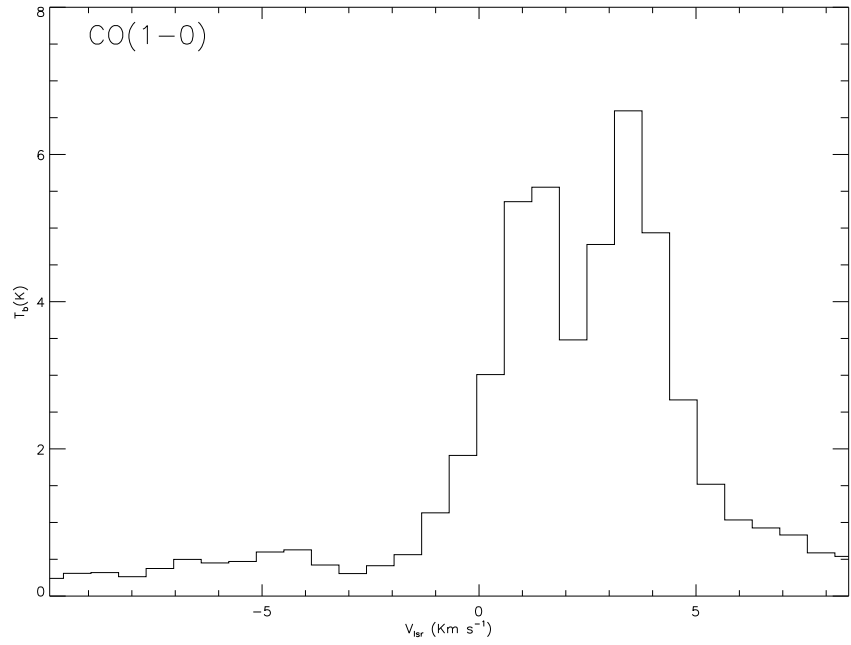}
\end{center}
\caption{The spectrum of PV Cep from CO(1$\rightarrow$0). The channel width is 0.62km sec$^{-1}$. The double peaks are consistent with an outflow emission. The emission drops at 2-3km sec$^{-1}$, which corresponds to the cloud velocity.}
\end{figure}

\clearpage
\begin{figure}
\begin{center}
\includegraphics[angle=-90,scale=0.45]{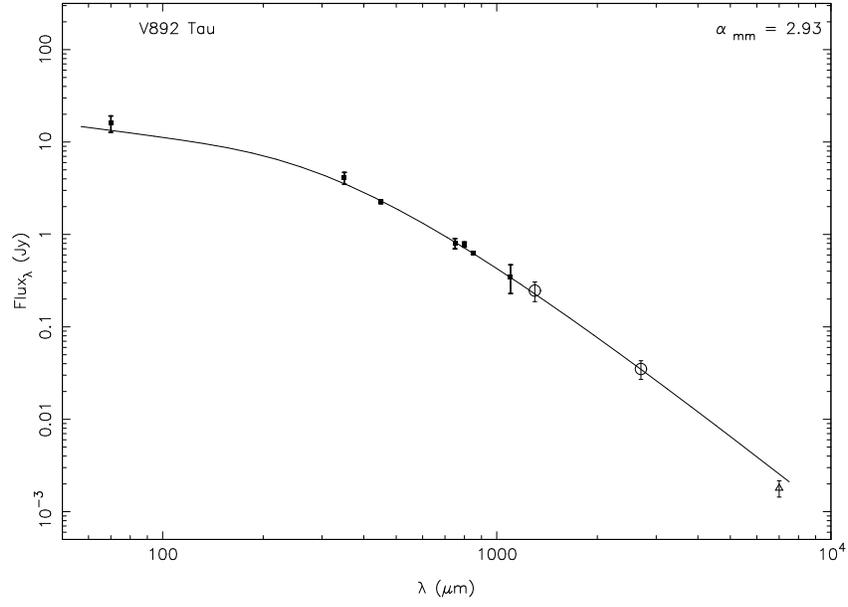}
\end{center}
\caption{SED of V892 Tau after correcting the fluxes from free-free contamination. The solid-line is a typical fit obtained for $\beta$=1, M$_D$=0.028\msun, $q$=0.8, and T$_{1AU}$= 550 K the corresponding $\chi^2_r \simeq$0.9. CARMA 1.3 and 2.7 mm data are shown with open circles. The millimeter spectral index $\alpha_{mm}$ is shown in the top right, and $F \propto \lambda^{-2.93}$.}
\end{figure}

\clearpage
\begin{figure}
\begin{center}
\includegraphics[angle=-90,scale=0.45]{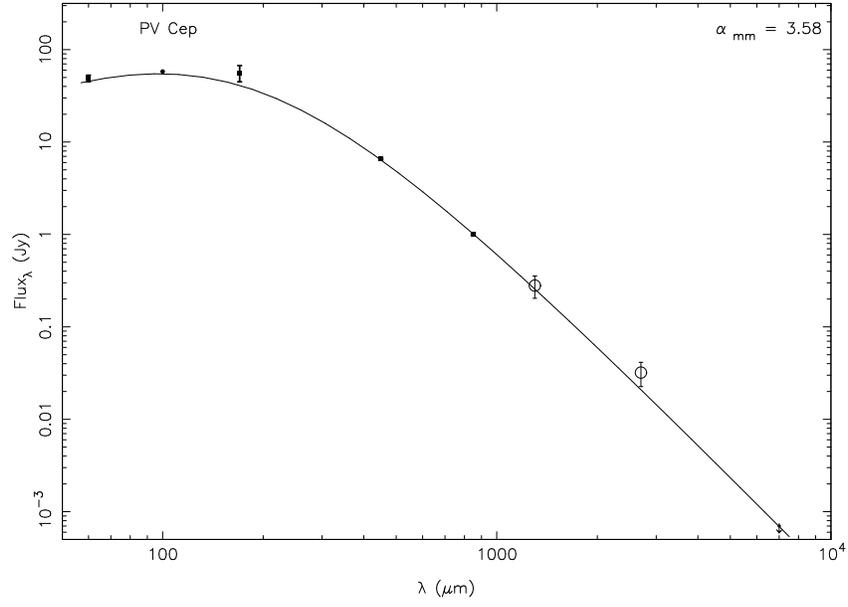}
\end{center}
\caption{SED of PV Cep after correcting the fluxes from free-free contamination. The solid-line is a typical fit obtained for $\beta$=1.7, M$_D$=0.72\msun, $q$=0.5, and T$_{1AU}$= 700 K the corresponding $\chi^2_r \simeq$0.8. The CARMA data at 1.3 and 2.7 mm data are shown with open circles. The VLA data at 7 mm is an upper limit. The millimeter spectral index $\alpha_{mm}$ is shown in the top right, and $F \propto \lambda^{-3.58}$.}

\end{figure}

\clearpage
\begin{deluxetable}{lcccccccc}
\tabletypesize{\scriptsize}
\tablecaption{Interferometric Observations.\label{tbl1}}
\tablewidth{0pt}
\tablehead{
\colhead{Object} & \colhead{RA} & \colhead{DEC} & \colhead{$\lambda$} &
\colhead{Array} & \colhead{Obs.} & \colhead{Beam} & \colhead{Beam} & \colhead{Phase} \\
\colhead{} & \colhead{(J2000)} & \colhead{(J2000)}  & \colhead{(mm)} & \colhead{Config.}  & \colhead{Date} & \colhead{FWHM($''$)} & \colhead{P.A.($^o$)} & \colhead{Calibrators}
}
\startdata
V892 Tau & 04 18 40.60 & +28 19 16.7 & 1.3 & CARMA C & 2007 Sept. 07 & 1.28 $\times$ 0.92 & -85 & 3C111 \\
... & ... & ... & 2.7 & CARMA B & 2008 Jan. 14 & 0.87 $\times$ 0.53 & -61 & 3C111\\
PV Cep & 20 45 53.96 & +67 57 38.9 & 1.3 & CARMA C & 2007 Sept. 19 & 1.07 $\times$ 0.74 & -59 & 1849+670   \\ 
...  & ... & ... & 2.7 & CARMA D & 2007 Jun. 06 &  6.4 $\times$ 4.8 & 65 & 1927+739\\
...  & ... & ... & 2.7 & CARMA B & 2008 Feb. 14 &  0.85 $\times$ 0.67 & 22 & 1927+739 \\
...  & ... & ... & 7 & VLA C & 2005 Jul. 25 & 0.65$\times$0.54 & -8.4 & 2006+644
\enddata

\end{deluxetable}


\begin{deluxetable}{lccccccc}
\tabletypesize{\scriptsize}
\tablecaption{Continuum Observations Results\label{tbl2}}
\tablewidth{0pt}
\tablehead{
\colhead{Object} & \colhead{$\lambda$} & \colhead{Size} & \colhead{incl.} & \colhead{P.A.} & \colhead{$R_{g}$} & \colhead{Total Flux} & \colhead{Disk} \\
\colhead{} & \colhead{(mm)} & \colhead{($''$)} & \colhead{($^o$)} & \colhead{($^o$)} & \colhead{(AU)} & \colhead{(mJy)} & \colhead{Resolved:} 
}
\startdata
V892 Tau & 1.3 & - & - & - & $<$ 170 & 251 & No \\
...  & 2.7 & 0.51$\times$0.27 & 59$\pm$6 & 64$\pm$5 & 93 & 39 & Yes \\
PV Cep & 1.3 &  0.68$\times$0.32 & 62$\pm$4 & -63$\pm$3 & 480 & 280 & Yes \\ 
...  & 2.7 & 0.64$\times$0.47 & 43$\pm$15 & 12$\pm$14 & 450 & 32 & Partially \\
... & 7 &  - & - & - & - & $<$0.76 & -

\enddata

\end{deluxetable}



\begin{deluxetable}{lccccccc}
\tabletypesize{\scriptsize}
\tablecaption{Disk properties deduced from SED modeling\label{tbl3}}
\tablewidth{0pt}
\tablehead{
\colhead{Object} & \colhead{M$_D$} & \colhead{$\beta$} & \colhead{T$_{1AU}$} & \colhead{q} & \colhead{$\chi^2_r$} \\
\colhead{} & \colhead{(\msun)} & \colhead{} & \colhead{(K)} & \colhead{} & \colhead{} 
}
\startdata
V892 Tau & 0.035$\pm$0.015 & 1.1$\pm$0.1 & 550$\pm$50 & 0.8 & $\lesssim$0.9\\ 
PV Cep & 0.76$\pm$0.04 & 1.75$\pm$0.05 & 700 &  0.5 & $\lesssim$0.9 

\enddata

\end{deluxetable}

\end{document}